\documentclass[a4paper]{jpconf}
\usepackage{graphicx}  

\begin{document}

%
\title{The behaviour of constrained caloric curves as ultimate signature of
a phase transition for hot nuclei}

%
\author{B. Borderie$^1$, S. Piantelli$^2$, M. F. Rivet$^1$, Ad. R.
Raduta$^3$,\\ E. Bonnet$^4$, R. Bougault$^5$, A. Chbihi$^4$, E.
Galichet$^{1,6}$,
D. Guinet$^7$,\\ Ph. Lautesse$^7$, N. Le Neindre$^5$, O. Lopez$^5$,
M. Parlog$^5$, E. Rosato$^8$,\\ R. Roy$^9$, G. Spadaccini$^8$, E. Vient$^5$,
M. Vigilante$^8$}

%
\address{
$^1$Institut de Physique Nucl\'eaire, CNRS/IN2P3, Univ. Paris-Sud 11, Orsay,
France\\
$^2$Sezione INFN, Sesto Fiorentino (Fi), Italy\\	 
$^3$National Institute for Physics and Nuclear Engineering, Bucharest-M\u{a}gurele,
Romania\\
$^4$GANIL, DSM-CEA/CNRS-IN2P3, Caen, France\\ 
$^5$LPC Caen, CNRS/IN2P3, ENSICAEN, Univ. de Caen, Caen, France\\	 
$^6$Conservatoire National des Arts et M\'etiers, Paris, France\\	 
$^7$Institut de Physique Nucl\'eaire, IN2P3-CNRS et Univ. Lyon I, Villeurbanne,
France\\
$^8$Dip. di Scienze Fisiche e Sezione INFN, Univ. Federico II, Napoli, Italy\\
$^9$Laboratoire de Physique Nucl\'eaire, Univ. Laval, Qu\'ebec, Canada}

\ead{borderie@ipno.in2p3.fr}

\begin{abstract}
Simulations based on experimental data obtained from multifragmenting quasi-fused nuclei
produced in central $^{129}$Xe + $^{nat}$Sn collisions have been
used to deduce event by
event freeze-out properties on the thermal excitation energy range
4-12 AMeV. From these properties and temperatures deduced from proton
transverse momentum fluctuations constrained caloric
curves have been built. At constant average volumes caloric curves
exhibit a monotonous behaviour whereas for constrained pressures
a backbending
is observed. Such results support the existence
of a first order phase transition for hot nuclei.
\end{abstract}

%
\section{Introduction}
One of the most important challenges of heavy-ion collisions at intermediate
energies is the identification and characterization of the nuclear
liquid-gas phase transition for hot nuclei, which was earlier theoretically
predicted for nuclear
matter~\cite{I46-Bor02}. In the last fifteen years a big effort
to accumulate experimental indications of the phase transition has been
made. Statistical mechanics for finite systems appeared as a key issue to
progress by proposing new first-order phase transition signatures related
to thermodynamic anomalies like negative microcanonical heat capacity and
bimodality of an order parameter~\cite{WCI06,Bor08,I72-Bon09}.
Correlated temperature and excitation energy measurements,
commonly termed caloric curves, were the first studied possible
signatures of phase transition. However in spite of the observation
of a plateau in some caloric curves, no decisive conclusion
could be extracted~\cite{I46-Bor02,Poc95,I8-Ma97,Nato02}. The reason is
that experimentally it is not possible
to explore the caloric curves at constant pressure or constant average
volume, which is required for an unambiguous phase transition signature.
Indeed, theoretical studies show that if many different caloric curves can
be generated depending on the path followed in the thermodynamical
landscape, constrained caloric curves exhibit different
behaviours in presence of a first order phase transition:
a monotonous evolution at constant average volume and a back
bending of curves at constrained pressures~\cite{Cho00,Fur06}. 
With the help of a simulation able to correctly reproduce
most of the experimental observables measured for
hot nuclei formed in central collisions (quasi-fused systems, QF, from
$^{129}Xe$+$^{nat}Sn$, 32-50 AMeV), event by event properties
at freeze-out were restored and used to build constrained
caloric curves.
The definition of pressure in the microcanonical ensemble
is presented in section 2.
Then, in section 3, simulations to recover
freeze-out properties of multifragmentation events~\cite{I58-Pia05,I66-Pia08}
are briefly described and constrained caloric
curves deduced are discussed. Section 4 is dedicated to
the description of the thermometer finally chosen and
to comparisons of results relative to
a caloric curve without any constraints.
Constrained caloric curves with temperatures including
quantum fluctuations are presented in section 5.
Section 6 is devoted to a discussion of the various results.
Conclusions are given in section 7.
\section{Pressure in the microcanonical ensemble}
Let us consider a gas of weakly interacting fragments 
(i.e. they interact only by Coulomb and excluded volume),
which corresponds to the freeze-out configuration.
Within a microcanonical ensemble, the statistical width of a configuration
$C$, defined by the mass, charge and internal excitation energy
of each of the constituting $M_C$ fragments,
writes
\begin{eqnarray}
\nonumber
W_C(A,Z,E,V) = \frac1{M_C!} \chi V^{M_C} \prod_{n=1}^{M_C}\left( 
\frac{\rho_n(\epsilon_n)}{h^3}(mA_n)^{3/2}\right)
\\ 
~ \frac{2\pi}{\Gamma(3/2(M_C-2))} ~ \frac{1}{\sqrt{({\rm det} I})}
~ \frac{(2 \pi K)^{3/2M_C-4}}{(mA)^{3/2}},
\label{eq:wc}
\end{eqnarray}
where $I$ is the moment of inertia,
$K$ is the thermal kinetic energy,
$V$ is the freeze-out volume and
 $\chi V^{M_C}$ stands for the free volume or, equivalently, accounts for
inter-fragment interaction in the hard-core idealization. 

The microcanonical equations of state are
\begin{eqnarray}
\nonumber
T=\left(\frac{\partial S}{\partial E}\right)^{-1}|_{V,A},\\
\nonumber
P/T=\left(\frac{\partial S}{\partial V}\right)|_{E,A},\\
-\mu/T=\left(\frac{\partial S}{\partial A}\right)|_{E,V}.
\end{eqnarray}

Taking now into account that $S=\ln Z=\ln \sum_C W_C$ 
and that $\partial W_C/\partial V=\left(M_C/V\right) W_C$, 
it comes out that
\begin{eqnarray}
\nonumber
P/T=
\left(\frac{\partial S}{\partial V}\right)&=&\frac1{\sum_C W_C} \sum_C 
\frac{\partial W_C}{\partial V}\\
&=&\frac1V \frac{\sum_C M_C W_C}{\sum_C  W_C}=\frac{<M_C>}{V}.
\end{eqnarray}

The microcanonical temperature is also easily deduced from its statistical
definition~\cite{Radut02}:
\begin{eqnarray}
\nonumber
T=\left(\frac{\partial S}{\partial E}\right)^{-1}&=&(\frac1{\sum_C W_C} \sum_C 
W_C(3/2M_C-5/2)/K)^{-1}\\
&=&<(3/2M_C-5/2)/K>^{-1}.
\end{eqnarray}
where $K$ is the total thermal kinetic energy of the system at freeze-out.

As $M_C$, the total multiplicity at freeze-out, is large, the pressure $P$ can
be well approximated by
\begin{eqnarray}
\nonumber
P=
\frac{2}{3}\frac{<K>}{V}.
\end{eqnarray}
\begin{figure}[h]
\centering
\includegraphics*[width=0.96\textwidth]
{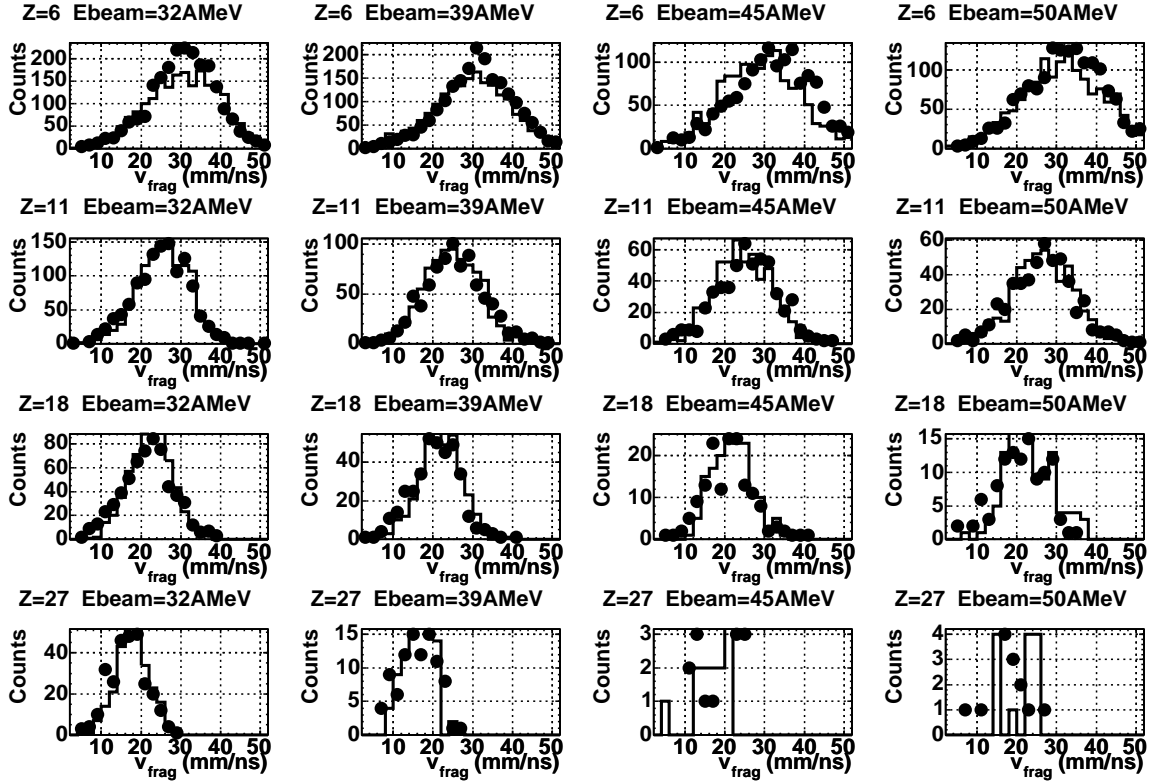}
\caption{ Comparison between the experimental velocity spectra (full points)
of fragments of a given charge and the simulated ones (histograms).
Each row refers to a different fragment charge: starting from the top
Z=6, Z=11, Z=18 and Z=27. Each column refers to a different beam
energy: starting from the left 32, 39, 45 and 50 AMeV.
From~\cite{I66-Pia08}.}
\label{fig1}
\end{figure}
\begin{figure}[h]
\centering
\includegraphics*[width=0.96\textwidth]
{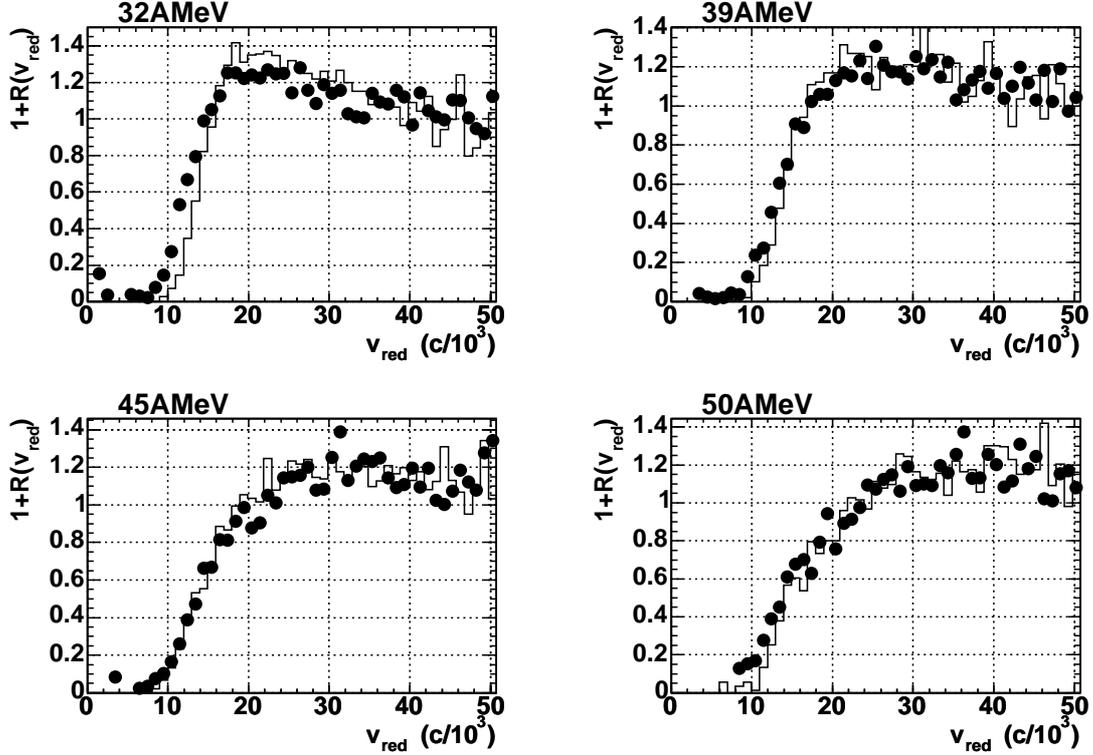}
\caption{Comparison betwwen the experimental (full points)
and simulated (histograms) reduced relative velocity correlation
functions for fragments. Each panel refers to a different
beam energy. From~\cite{I66-Pia08}.}
\label{fig2}
\end{figure}
\section{Event by event freeze-out properties}
\begin{figure}[h]
\centering
\includegraphics*[width=0.96\textwidth]
{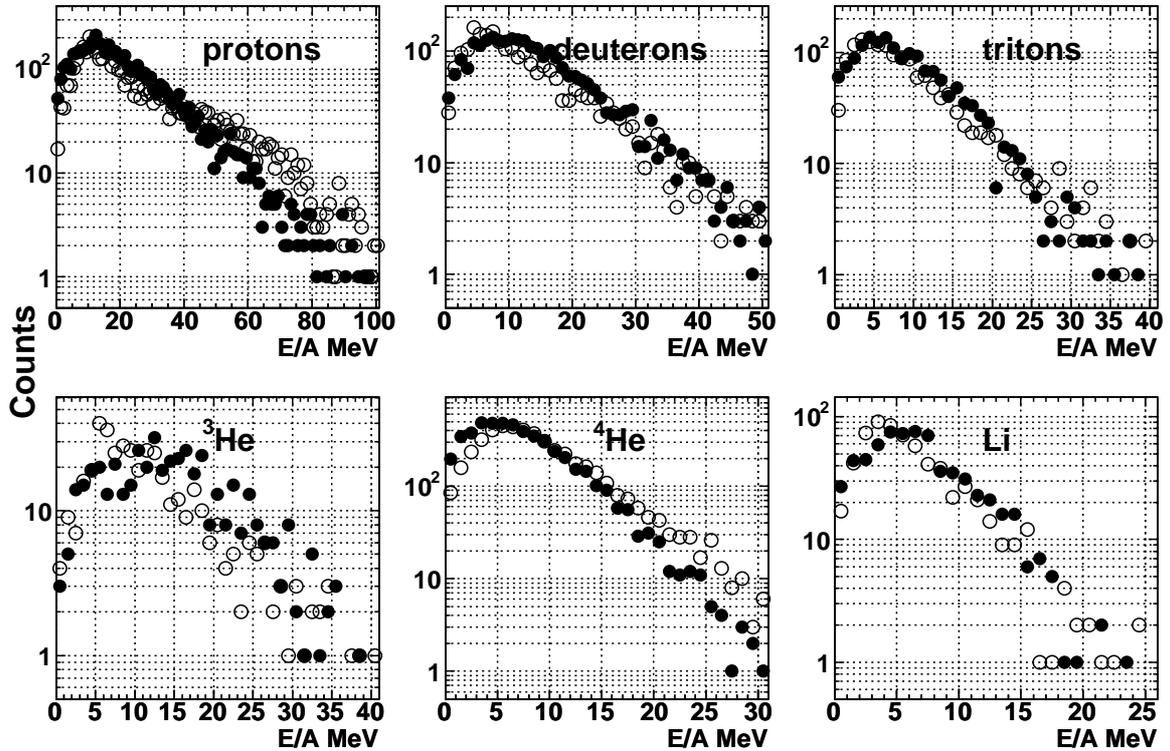}
\caption{Centre of mass energy spectra for protons, deuterons, tritons,
$^3$He, $^4$He and $^7$Li for the reaction at 39 AMeV.
Open points are the simulated data, while full points are the
experimental ones. From~\cite{I66-Pia08}.}
\label{fig3}
\end{figure}

Starting from all the available asymptotic experimental information
(charged particle energy spectra, average and standard deviation of fragment
velocity spectra and calorimetry) of selected QF
sources produced in central $^{129}$Xe+$^{nat}$Sn collisions which undergo
multifragmentation, a simulation was performed to reconstruct freeze-out
properties event by event~\cite{I58-Pia05,I66-Pia08}. 
The method requires data with a very high degree
of completeness (total detected charge $\geq$93\% of the total
charge of the system), which is crucial for a good estimate of Coulomb energy.
QF sources are reconstructed, event by event, in the reaction centre of mass,
from all the fragments and twice the charged particles emitted in the range
$60-120^{\circ}$ in order to exclude a major part of pre-equilibrium emission.
Dressed excited fragments and particles at
freeze-out are described by spheres at normal density.
Then the excited fragments subsequently deexcite while flying away.
Four free parameters
are used to recover the data at each incident energy: the percentage
of measured particles which were
evaporated from primary fragments, the collective radial energy, a minimum
distance between the surface of products at freeze-out and a limiting
temperature for fragments.
The limiting temperature,
related to the vanishing of level density for fragments~\cite{Koo87},
was mandatory to reproduce the observed widths of fragment velocities.
Indeed, Coulomb repulsion plus collective energy 
plus thermal kinetic energy (directed at random) plus spreading due to fragment
decays are responsible for about 60-70\% of the observed widths.
By introducing a limiting temperature for fragments, thermal kinetic
increases, due to energy conservation, which produces the missing
percentage for the widths of final velocity distributions.
The agreement between experimental and simulated velocity/energy spectra for
fragments and for the
different beam energies is quite remarkable 
(see figure \ref{fig1}).
Relative velocities between fragment pairs were also compared
through reduced relative velocity correlation 
functions~\cite{Kim92,I57-Tab05}
(see figure \ref{fig2}).
Again a good agreement is obtained between experimental data and
simulations, which
indicates that the retained method (freeze-out topology built up
at random) and parameters are sufficiently relevant
to correctly describe the freeze-out configurations, including volumes.
Finally one can also note that the agreement between experimental and simulated 
energy spectra for protons and
alpha-particles (see an example in figure \ref{fig3}) is not very good at
high energy. This
comes from the fact that we have chosen (to limit the number of parameters of
the simulation) a single value for the percentage of all measured
particles which were evaporated from primary fragments. We shall
see in the following section how to correct for temperature measurements
derived from protons.

From the simulations it is then possible to recover, event by event, the different
quantities needed to build constrained caloric curves, namely
the thermal excitation energy of QF hot nuclei, $E^*$ (total excitation minus
collective energy) with an estimated systematic error of around 1 AMeV,
the kinetic temperature $T_{kin}$ at freeze-out,
the freeze-out volume $V$ (see envelopes of figure 8 from~\cite{I66-Pia08}) and
the total thermal kinetic energy at freeze-out $K$.
In simulations, Maxwell-Boltzmann distributions
are used to reproduce the thermal kinetic properties at freeze-out
and consequently
the deduced temperatures, $T_{kin}$, are classical.

\begin{figure}[h]
\centering
\includegraphics*[width=0.48\textwidth]
{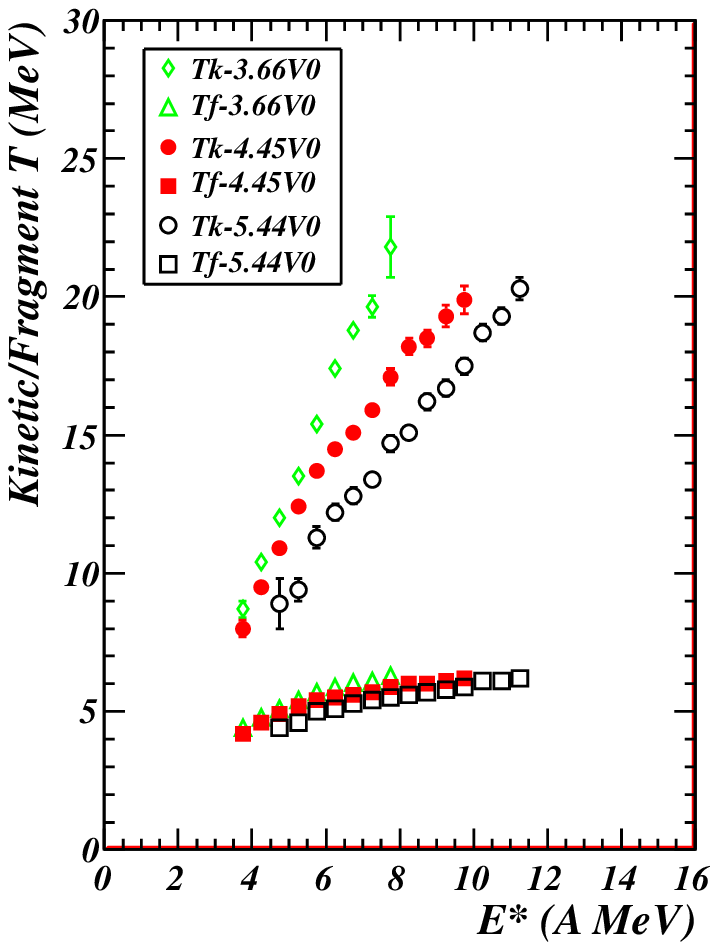}
\includegraphics*[width=0.48\textwidth]
{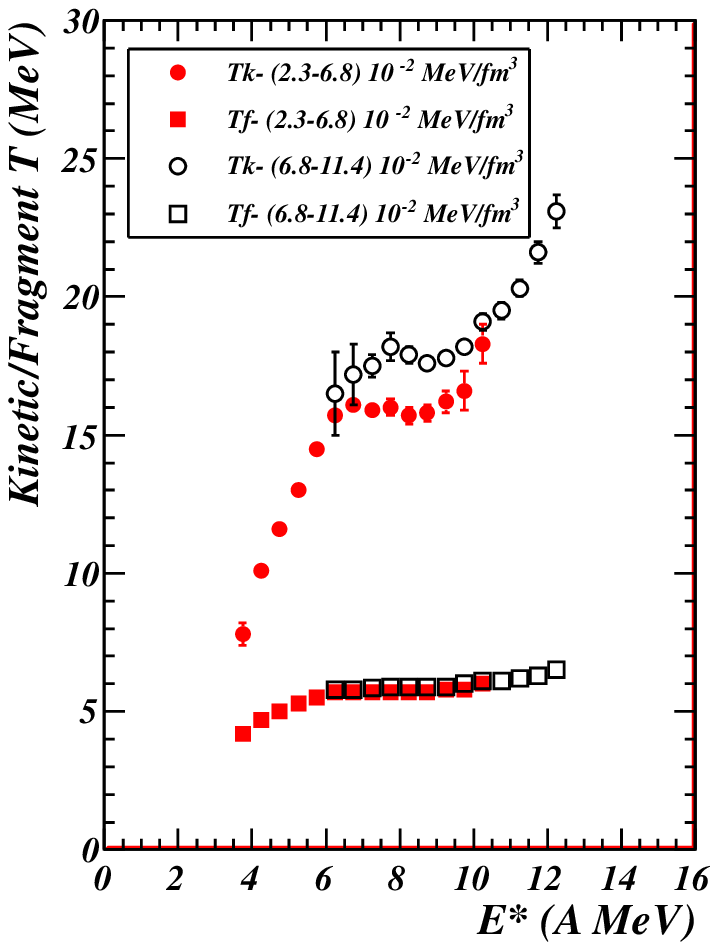}
\caption{Caloric curves (kinetic temperature versus thermal
excitation energy) constrained at average volumes (left)
and for selected ranges of pressure (right) and the corresponding internal
temperatures for fragments. Error bars correspond to statistical errors.
From~\cite{BBIWM11}.}
\label{fig4}
\end{figure}

Constrained caloric curves, which correspond to correlated values of
$E^*$ and $T_{kin}$ have been derived for QF hot nuclei with Z restricted
to the range 80-100, which corresponds to the A domain 194-238, in order to
reduce any possible effect of mass variation on caloric curves~\cite{Nato02}
(see figure~\ref{fig4}).
Curves for internal fragment temperatures are also shown in the figure.
For different average freeze-out volumes expressed as a function of
$V_0$, 
the volume of the QF nuclei at normal density,
a monotonous behaviour of caloric curves is observed
as theoretically expected.
The caloric curves when pressure has been constrained 
exhibit a backbending and moreover the qualitative evolution of curves
with increasing pressure exactly corresponds to what is theoretically
predicted with a microcanonical lattice gas model~\cite{Cho00}.
However by extrapolating at higher pressures, 
one could infer a critical temperature around 20 MeV. Such a value is
within the range calculated for infinite nuclear matter
whereas a lower value is expected for nuclei in relation with surface
and Coulomb effects~\cite{I46-Bor02}. At that point, one may
wonder if $T_{kin}$ is a relevant thermometer. We shall see, in what follows, that 
the final choice was to use, for protons thermally emitted at freeze-out, a new
thermometer recently proposed for which quantum effects can be included.

\section{Temperatures from proton transverse momentum fluctuations}
Very recently a new method for measuring the temperature of hot nuclei was
proposed~\cite{Wue2010,Zhe11}. It is based on momentum fluctuations of emitted
particles like protons in the centre of mass frame of the fragmenting nuclei. On the
classical side, assuming a Maxwell-Boltzmann distribution of the momentum
yields, the temperature T is deduced from the quadrupole momentum fluctuations
defined in a direction transverse to the beam axis:\\
$\sigma^2$ = $<Q_{xy}^2>$ - $<Q_{xy}>^2$ = $4m^2T^2$\\
with $Q_{xy}$ = $P_{x}^2$ - $P_{y}^2$;
m and P are the mass and linear momentum of emitted particles.
Taking into account the quantum nature of particles, a correction
$F_{QC}$
related to a Fermi-Dirac distribution was also proposed~\cite{Zhe11,Zhe12}.\\
In that case $\sigma^2$ = $4m^2T^2$ $F_{QC}$ where $F_{QC}$ =
$0.2(T/\epsilon_f)^{-1.71}$ + 1;\\ $\epsilon_f$ = 36 $(\rho/\rho_0)^{2/3}$
is the Fermi energy of nuclear matter at density $\rho$ and
$\rho_0$ corresponds to normal density.

\begin{figure} [h]
\begin{minipage} {17pc}
\vspace{1.0pc}
\includegraphics[width=17pc]{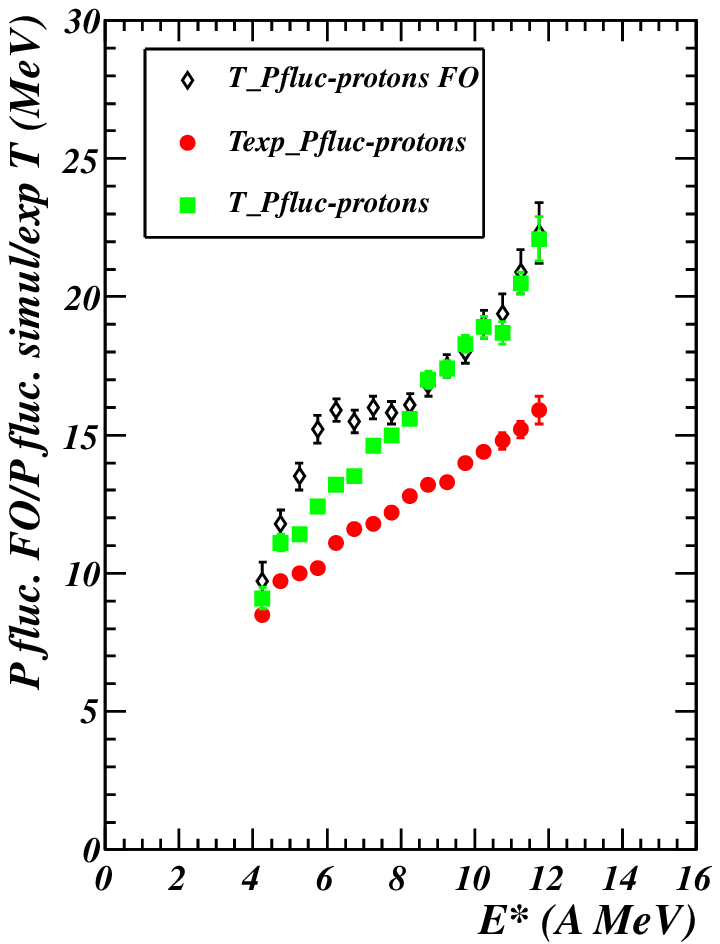}
\caption{\label{fig5}
Caloric curves (classical temperature from proton transverse momentum
fluctuations versus thermal excitation energy) for protons (simulation)
thermally emitted at freeze-out (open diamonds), for protons (simulation)
after the secondary decay stage (full squares),
and from protons experimentally measured (full points). Error bars
correspond to statistical errors}
\end{minipage}\hspace{2pc}%
\begin{minipage} {17pc}
\includegraphics[width=17pc]{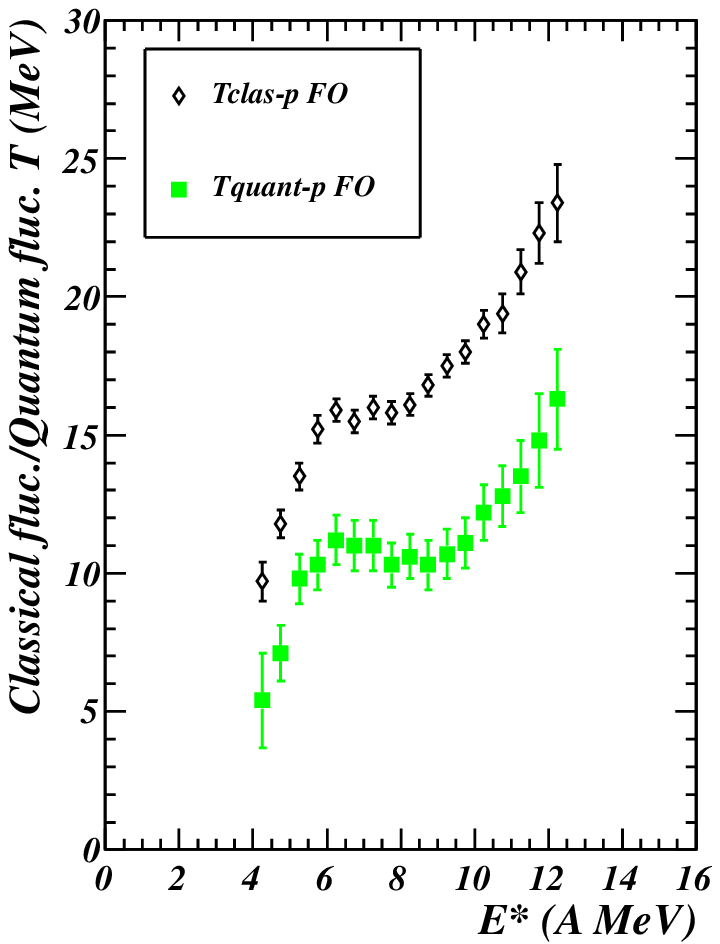}
\caption{\label{fig6} Caloric curves: classical temperature (open diamonds)/
quantum corrected temperature (full squares) 
from proton transverse momentum
fluctuations versus thermal excitation energy. Protons (simulation)
are thermally emitted at freeze-out. Error bars include statistical and
systematic errors.}
\end{minipage}
\end{figure}

Before using the thermometer with protons to build constrained caloric curves, it was
important to verify several things. With the classical simulation
(freeze-out and asymptotic proton momenta) , it
is possible to test the agreement with the proposed classical thermometer.
Moreover the effects of secondary decays on temperature measurements can be
estimated. Figure~\ref{fig5} shows different caloric curves without constraints.
Note that the selection in Z and A of hot nuclei is the same as in the
previous section. It was verified that, within statistical error bars,
at a given thermal excitation energy transverse momentum fluctation values
are the same for our selection or
by selecting only a single (A and Z) hot nucleus. 
Open diamonds refer to classical temperatures calculated from momentum fluctuations
for protons thermally emitted at freeze-out. Within the statistical error
bars they perfectly superimposed on $T_{kin}$ values
(see figure 2 of~\cite{BBIWM11}), which just verifies
that Maxwell-Boltzmann distribution was correctly implemented in the
simulation. Full squares correspond to classical temperatures calculated from
momentum fluctuations for protons after the secondary decay stage. We note
that the caloric curve is distorted, which means that it is not possible to
use experimental data from protons to measure temperatures. Moreover, in
this case, quantum corrections for temperatures can not be made because protons
are emitted at different stages of deexcitation with different Fermi energy values.
In figure~\ref{fig5} classical temperatures calculated from experimental proton
data are also shown (full points). As for temperatures calculated from asymptotic proton
data of simulations, a monotonous behaviour of the caloric curve is observed.
One also note the differences between the two sets of temperature values,
which are related to the
fact that, as indicated in the previous section, simulations do not describe
experimental proton energy spectra very well. Those temperature differences
will be used to correct classical temperatures derived from simulated protons
at freeze-out.

It finally appears that the only way to extract temperatures from proton transverse momentum
fluctuations taking into account quantum effects is to use protons thermally
emitted at freeze-out. In that case classical temperature values from simulations
must be extracted and corrected and then, quantum corrections applied, which
needs Fermi energy values. Those values can be estimated from semi-classical
calculations (Xn+Sn at 32 AMeV and Sn+Sn at 50 AMeV)~\cite{FRA01,RIZ07}:
protons thermally emitted at freeze-out at time around 100-120 fm/c after the
beginning of collisions come from a low density uniform source. For the two
incident energies low densities around $\rho\sim0.4\rho_0$ are calculated
which corresponds to $\epsilon_f\sim$ 20 MeV. We have introduced a systematic
error of $\pm$ 0.1$\rho_0$ for the calculation of $\epsilon_f$ and consequently
a systematic error for ``quantum'' temperatures of $\pm$ 0.6-0.5 MeV on the
considered thermal excitation energy range. Figure~\ref{fig6} shows the final
caloric curve with temperatures from quantum fluctuations (full squares).
It exhibits a plateau around a temperature of 10-11 MeV on the $E^*$ range
5-10 AMeV.
For comparison the caloric curve with classical temperatures derived from
the simulation is also shown (open diamonds).
  
\section{Constrained caloric curves with quantum temperatures }
\begin{figure}[h]
\centering
\includegraphics*[width=0.48\textwidth]
{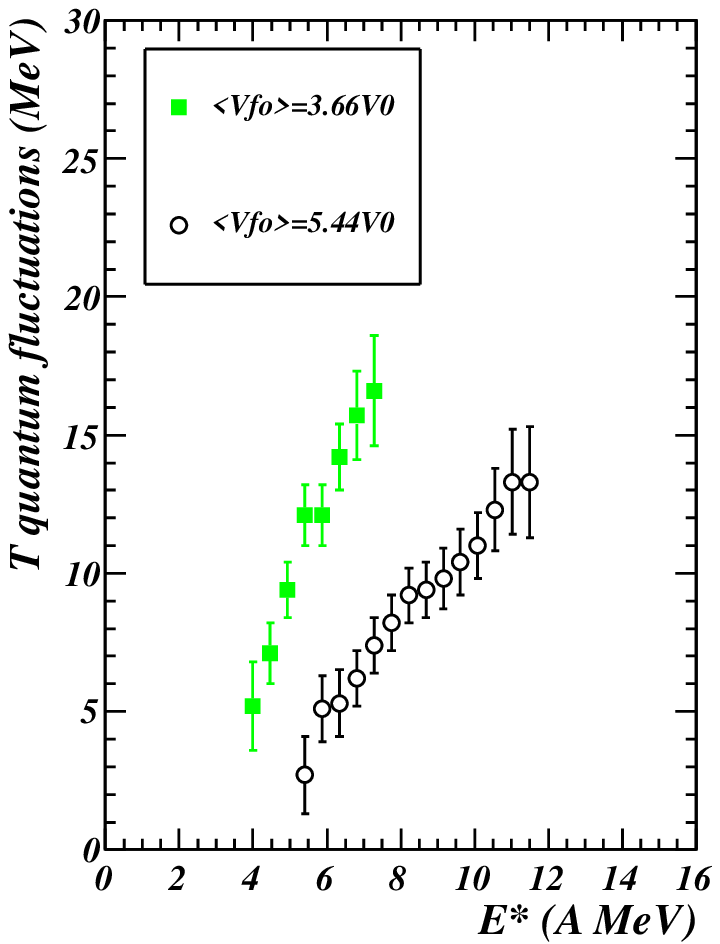}
\includegraphics*[width=0.48\textwidth]
{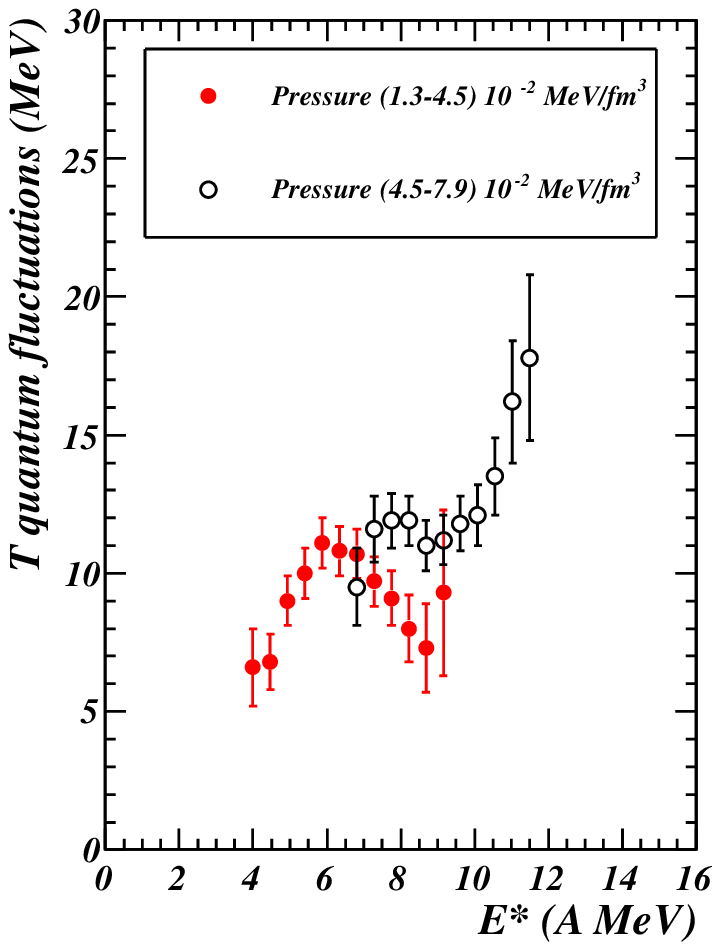}
\caption{Caloric curves (quantum corrected temperature versus thermal
excitation energy) constrained at average volumes (left)
and for selected ranges of pressure (right).
Error bars include statistical and
systematic errors.}
\label{fig7}
\end{figure}

Constrained caloric curves, which correspond to correlated values of 
$E^*$ and quantum corrected temperatures have been determined.
Accordingly $E^*$ values, which were initially derived from experimental
calorimetry including estimated correction for neutrons
(see~\cite{I58-Pia05}),
have been corrected a posteriori using quantum temperatures at freeze-out.
Pressure values were also corrected using quantum temperatures.
In figure~\ref{fig7} (left) we have constructed caloric curves for two different
average freeze-out volumes corresponding to the ranges
3.0-4.0$V_0$ and 5.0-6.0$V_0$ where $V_0$ correspond to 
the volume of the QF nuclei at normal density.
Again as theoretically expected a monotonous behaviour of caloric curves is observed.
Figure~\ref{fig7} (right) shows the caloric curves when pressure has been constrained
within two domains: 1.3-4.5 and 4.5-7.9 $10^{-2}$ MeVfm$^{-3}$.
Backbending are clearly seen especially for the lower pressure range. 
For higher pressures the backbending of the caloric curve is largely reduced
and one can estimate that the critical temperature is around 12-13 MeV for
the selected hot nuclei. 
\begin{figure}[h]
\centering
\includegraphics*[width=0.48\textwidth]
{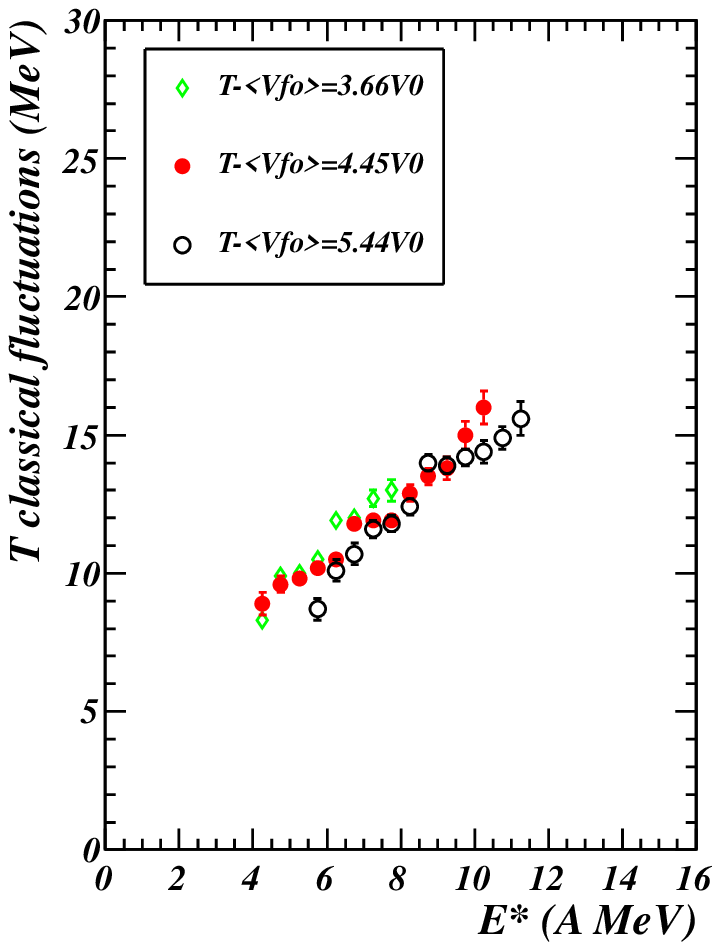}
\includegraphics*[width=0.48\textwidth]
{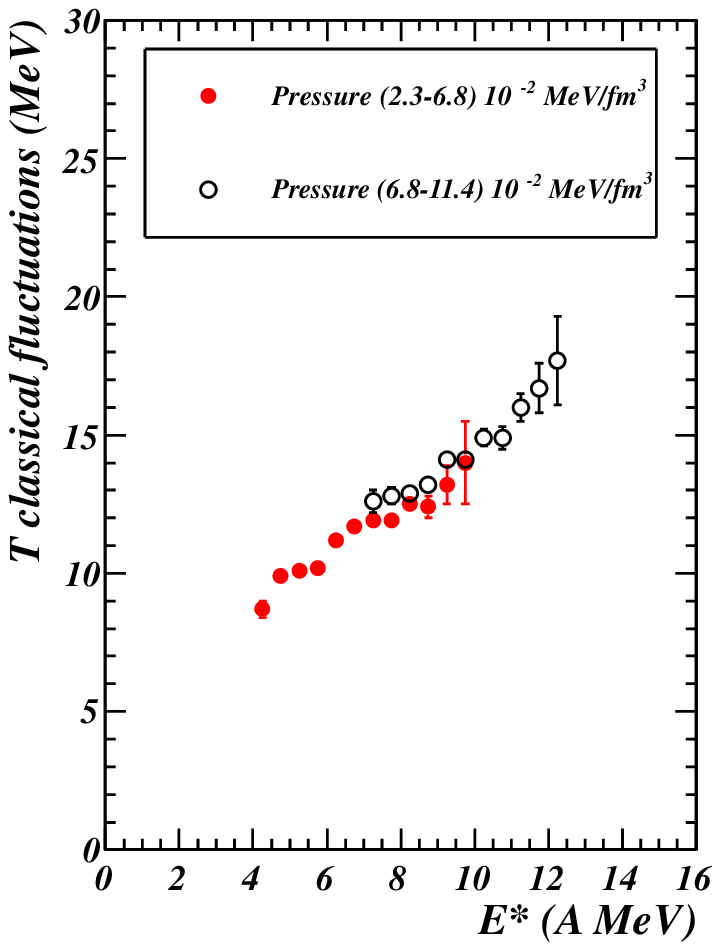}
\caption{Caloric curves (classical temperature derived from experimental
proton data versus thermal
excitation energy) constrained at average volumes (left)
and for selected ranges of pressure (right).
Error bars correspond to statistical errors.}
\label{fig8}
\end{figure}
\section{Discussion}
First of all, as far as internal temperature for fragments are concerned (see
figure~\ref{fig4}), one can observe that they perfectly agree, for our A range,
with those calculated with the well known ``He/Li thermometer''
used in~\cite{Poc95,Nato02} keeping the proposed prefactor 16. Note that
similar temperature values are also obtained from our data (see figure 2
from~\cite{BBIWM11}). This is a strong indication that the He/Li thermometer 
mainly reflects the internal temperature of fragments in relation with the
strong secondary emission of alpha-particles and Li isotopes.
As indicated in~\cite{Nato02} those plateau temperatures can be interpreted
as representing the limiting temperatures resulting from Coulomb instabilities
of heated nuclei which has long been predicted~\cite{Lev85}.
Indeed, for thermally equilibrated QF hot nuclei one expects internal temperature
for fragments equal to the limiting temperature of the fragmenting
system. As a direct consequence
internal fragment temperatures must reflect the evolution of limiting
temperature with A of hot nuclei, which is indeed experimentally 
observed~\cite{Nato02}. 

For a finite piece of nuclear matter like a hot nucleus the microcanonical
ensemble is the most relevant ensemble and the kinetic/microcanonical
temperature is the relevant parameter to build caloric curves and 
deduce information on a possible phase transition. On the other hand,
one may also 
recall that dispersions in fragment transverse momentum spectra generated in
projectile fragmentation were succesfully explained by adding the
Fermi momenta of individual nucleons in fragments~\cite{GOL74},
which is a consequence of the quantum nature of nucleons. For
multifragmentation reactions a similar approach was first proposed
in~\cite{Bau95} to qualitatively explain the different temperatures
obtained from various observables (kinetic energies, level populations,
isotope ratios). Thus, the use of the thermometer recently proposed based on
quantum fluctuations~\cite{Zhe11} was a good opportunity to better
investigate the behaviour of caloric curves especially under constraints in
volume and pressure.
The behaviour predicted by theoretical works for a first order phase transition
is observed  and confirms results obtained for other signals: negative
microcanonical heat capacity and bimodality of an order parameter.

The last point that we want to discuss concerns information which can be
directly deduced from experimental data like proton transverse momentum
fluctuations. In that case as previously indicated Fermi energies can not be
estimated and only classical temperatures can be calculated.
From simulations we have seen previously that secondary decays distorted
caloric curves.
Results displayed in figure~\ref{fig8} fully confirm this observation:
constrained caloric curves built with classical temperatures derived from
experimental proton momenta also exhibit a monotonous behaviour and do not show 
any dependence upon average volumes or pressures, which prevent any direct
information.

\section{Conclusions}
Several caloric curves have been derived for hot nuclei from quasi-fused
systems using a new thermometer based on proton transverse momentum fluctuations
including quantum effects. The unconstrained caloric curve exhibits a plateau at
a temperature of around 10-11 MeV on the thermal excitation energy range
5-10 AMeV. For constrained caloric curves (volume and pressure) we observe
what is expected for a first order phase transition for finite systems in
the microcanonical ensemble, namely a monotonous behaviour at constant
average volumes and backbending for constrained pressures.
After the observation of negative microcanonical heat capacity and bimodality
the behaviour of caloric curves is the ultimate signature of a
first order phase transition for hot nuclei. 

The exit, at high excitation, of the spinodal
region and of the coexistence region around respectively 8 and 10 AMeV are
in good agreement (within error bars) with spinodal~\cite{Tab03} and
bimodality~\cite{I72-Bon09} signals. Note also that 10 AMeV corresponds
to first experimental indications for the onset of
vaporization~\cite{Tsa93,Bor96,Bor99,Pie00}.

Finally, one can infer from the
present studies that the temperatures obtained with
the He/Li thermometer, which was largely used in the past, seem
mainly reflect the internal temperatures of fragments
in the excitation energy range 5-10 AMeV.

\ack
The authors are indebted to A.~Bonasera for private information and
discussions.é
\section*{References}



\end{document}